 \documentclass[final,5p,times,twocolumn]{elsarticle}
\usepackage{amsmath}
\usepackage{slashed}

\usepackage{amssymb}

\newcommand{\sfrac}[2]{{\textstyle\frac{#1}{#2}}}

\newcommand{\ihalf}{\sfrac{i}{2}}

\newcommand{\alg}[1]{\mathfrak{#1}}

\journal{Physics Letters B}

\begin{document}

\begin{frontmatter}



\title{AdS/dCFT one-point functions of the $S\!U(3)$  sector  }

\author{Marius de Leeuw, Charlotte Kristjansen, and Stefano Mori}
\address[label2]{Niels Bohr Institute, Copenhagen University,\\ Blegdamsvej 17, 2100 Copenhagen \O, Denmark}



\begin{abstract}
We propose a closed formula for the tree-level one-point functions of non-protected operators belonging to an $S\!U(3)$ sub-sector
of the defect CFT dual to the D3-D5 probe brane system with background gauge field flux, $k$, valid for $k=2$. The formula passes a number of non-trivial
analytical and numerical tests.  Our proposal is based on  expressing the one-point functions  as an overlap between a Bethe eigenstate of the $S\!U(3)$ spin chain and a certain matrix product state, deriving various  factorization properties of the Gaudin norm and performing explicit computations for shorter spin chains. 
As its $S\!U(2)$ counterpart, the one-point function formula for the  $S\!U(3)$ sub-sector  is of determinant type.  
We discuss the the differences with the $S\!U(2)$ case and the challenges in extending the present formula beyond $k=2$. 

\end{abstract}


\end{frontmatter}



\section{Introduction}
The integrable structure of N=4 SYM~\cite{Beisert:2010jr} 
has recently shown its power in the calculation of one-point functions in a
certain defect version of the theory which is holographically dual to the D3-D5 probe brane system with background
gauge field flux~\cite{Karch:2000gx,DeWolfe:2001pq,Constable:1999ac}.
Using the tools of integrability a closed formula of determinant form, valid for the 
tree-level one-point function of any non-protected operator
from the $S\!U(2)$-sector and for any value of the string theory background gauge field flux, $k$, was derived~\cite{deLeeuw:2015hxa,
Buhl-Mortensen:2015gfd}. 
The formula revealed interesting connections 
to recent work in condensed matter physics~\cite{Brockmann2014b,Brockmann2014,Brockmann2014a}. A burning question is, of course, whether the integrable structure
allows one to extend the closed formula for one-point functions to other sectors than the $SU(2)$ sub-sector, to higher
loop orders and to other set-ups involving defects. In reference~\cite{Buhl-Mortensen:2016pxs} the calculation of loop
corrections to one-point functions was initiated. Here, we 
take the first step in the direction of moving on to other sectors by presenting a closed formula for the tree-level one-point functions of non-protected operators for $k=2$ in the $S\!U(3)$ sector which is a closed sub-sector at one-loop order. In the 
$S\!U(2)$ case the $k=2$ formula provided the starting point of a recursive relation, based on the transfer matrix
of the integrable Heisenberg spin chain, which gave access to the one-point function for
any value of $k$. Earlier, tree-level one-point functions of protected operators (chiral primaries) involving all 
six scalar fields of ${\cal N}=4$ SYM were evaluated  both for the present dCFT dual to the D3-D5 probe brane system~\cite{Nagasaki:2012re} and for two instances of a dCFT dual to a D3-D7 probe brane system~\cite{Kristjansen:2012tn}. For this type of computations the tools of integrability are, however, not needed.

We start by outlining the definition of the defect CFT in section~\ref{holography}. Subsequently, in section~\ref{SU(2)}
we revisit the $S\!U(2)$ sub-sector and express the one-point functions for $k=2$ in a form which suggests a generalisation to  $S\!U(3)$ which we treat in section~\ref{SU(3)}. 
We end with a discussion and conclusion in section~\ref{Conclusion}.

\section{The holographic set-up and the defect CFT\label{holography}}

By considering a probe D5 brane with geometry $AdS_4\times S^2$  embedded in the usual $AdS_5\times S^5$ background
one can engineer a system whose dual is a
defect version of ${\cal N}=4$ SYM~\cite{Karch:2000gx}.
More precisely, the field theory consists of a co-dimension one defect placed at 
$z=0$ which is the home of a hyper multiplet of fundamental fields which have self interactions as well as interactions with
the bulk ${\cal N}=4$ SYM fields~\cite{DeWolfe:2001pq}.
If one furthermore arranges that $k$ of the usual $N$ D3 branes get dissolved into the D5 brane by
allowing a background gauge field to have flux $k$ through the $S^2$
one can arrive at a situation where the gauge-group of ${\cal N}=4$ SYM
is $S\!U(N-k)$ on one side of the defect and $S\!U(N)$ on the other. In the string theory language the dissolution of the
$k$ D3 branes into the D5 brane is described by the fuzzy funnel solution which in the field theory picture implies that 
three of the six scalar fields of ${\cal N}=4 $ SYM acquire a non-vanishing vacuum expectation value on one side of 
the defect~\cite{Constable:1999ac}.
More precisely
\begin{eqnarray}\label{Phiclass}
\!\!\!\!\!\!\!\Phi _i^{\rm cl}& =& -
 \frac{1}{z}\,\begin{pmatrix}
  \left(t_i\right)_{k\times k} & 0_{k\times (N-k)}\\ 
  0_{(N-k)\times k} & 0_{(N-k)\times (N-k)} \\ 
 \end{pmatrix},~i=1,2,3, \hspace{0.2cm} z>0\\
 \!\!\!\!\!\!\!\Phi ^{\rm cl}_i&=&0,~ i=4,5,6,
\end{eqnarray}
where the three $k\times k$ matrices $t_i$  constitute a $k$-dimensional unitary, irreducible representation of $\mathfrak{su}(2)$,
in particular
\begin{equation}\label{eq:su2relations}
 \left[t_i,t_j\right]=i\varepsilon _{ijk}t_k.
\end{equation}
For $z<0$ all classical fields are vanishing.
Given the vevs in eqn.~(\ref{Phiclass}) it is clear that operators 
constructed from fields of the type $\Phi_1, \Phi_2$ and $\Phi_3$ will have non-vanishing one-point functions already at tree-level. Applying the arguments of Cardy~\cite{Cardy:1984bb} one gets that one-point functions in the present  
CFT with a defect at $z=0$ 
are constrained to take the form
\begin{equation}
\langle {\cal O}_{\Delta}\rangle=\frac{C}{z^{\Delta}},
\end{equation}
where $C$ is a constant and $\Delta$ denotes the scaling dimension of the conformal operator ${\cal O}_{\Delta}$ in the theory without defect. It is well known that at one-loop order conformal single trace operators built from scalar fields are characterized as being Bethe eigenstates of an integrable  $S\!O(6)$ spin chain~\cite{Minahan:2002ve}.  This fact makes it possible to write
the one-point functions in the scalar sector as an overlap between a (unit normalized) Bethe eigenstate and a certain so-called matrix product
state~\cite{deLeeuw:2015hxa}.
\begin{equation}\label{genericCso6}
 C=
 \left(\frac{8\pi ^2}{\lambda }\right)^{\frac{L}{2}}L^{-\frac{1}{2}}
 \,C_k, \hspace{0.5cm} C_k=
 \frac{\left\langle {\rm MPS}_k\,\right.\!\!\left|\vphantom{{\rm MPS}_k}\Psi \right\rangle}{\left\langle \Psi \right.\!\!\left|\Psi  \right\rangle^{\frac{1}{2}}}.
\end{equation}
Here the pre-factor is a normalization factor ensuring the canonical normalization of the two-point functions of ${\cal N}=4$ SYM,
$L$ denotes the number of fields in the single trace operators, $\lambda$ is the 't Hooft coupling constant, 
$|\Psi \rangle$ the Bethe Eigenstate and $|\mbox{MPS}_k\rangle$ the matrix product state (to be detailed below) corresponding to the
$k$-dimensional irreducible representation of $S\!U(2)$.

\section{The $SU(2)$ sub-sector\label{SU(2)}}
An $S\!U(2)$ sub-sector of ${\cal N}=4$ SYM consists of operators constructed from two types of complex scalar fields each 
built out of two of the theory's real scalar fields. In the defect set-up described above an interesting $S\!U(2)$-sub-sector can
be constructed f.inst.\ as follows
\begin{align}
&Z=\Phi _1+i\Phi _4,
&&Y=\Phi _2+i\Phi _5.
\end{align}
The one-loop conformal operators of this sub-sector can be mapped to  Bethe eigenstates of the Heisenberg spin chain  \cite{Minahan:2002ve}. By exploiting the integrability of the Heisenberg spin chain, it was possible to derive a closed expression of
determinant type for $C_k$ valid for any operator of the $S\!U(2)$ sub-sector and for any value of $k$~\cite{deLeeuw:2015hxa,Buhl-Mortensen:2015gfd}.  A key role was played by the formula for $k=2$ as it turned out that $C_k$ for all higher
values of $k$ could be recursively related to this one. In the present letter we will present a similar closed formula
for $C_2$ for operators from the $S\!U(3)$ sector. To set the scene for that and to motivate our formula we shall start by
recapitulating the $S\!U(2)$ results from~\cite{deLeeuw:2015hxa} while slightly modifying the formulation.

The Heisenberg spin chain whose eigenstates correspond to the conformal operators is described by the Hamiltonian
\begin{equation}{\label{Hamiltonian}}
H=\sum_{l=1}^L 1-P_{l,l+1},
\end{equation}
where $L$ is the length of the chain (equal to the number of fields in the single trace operator) and $P$ is the permutation
operator. To each site of this spin chain is associated a Hilbert space $\mathbb{C}^2$ with basis vectors 
 $|e_{1}\rangle, |e_{2}\rangle $ corresponding to the fundamental representation of $S\!U(2)$ (spin up and spin down).
In the mapping of a single trace operator onto a spin chain state the field $Z$ is mapped to   $|e_{1}\rangle$ and the
field $Y$  to $|e_{2}\rangle$.
For convenience, let us introduce the matrix unities $E^i_j := |e_j\rangle\langle e_i| $. Furthermore, we denote the matrix unity acting on the $n$-th site of the spin chain  by $(E^i_j)_n$. As the vacuum of the spin chain we choose the state 
 $|0\rangle_L = |\!\uparrow\ldots\uparrow\rangle_L$. A Bethe state with $M$ excitations (spin down) can then be written
 as 
\begin{align}\label{eq:BSsu2}
|\{u_i\}\rangle :=\mathcal{N}  \sum_{1\leq m_1<\ldots <m_M \leq L}
\left[
\sum_{\sigma\in \mathcal{S}_M}
 \mathbb{S}_\sigma \prod_{r=1}^M \left[\frac{u_{\sigma_r}+\ihalf}{u_{\sigma_r}-\ihalf}\right]^{m_r} 
\right]
\prod_{s=1}^M  (E^1_2)_{m_s} |0\rangle,
\end{align}
where $\mathcal{N}$ is a normalization factor, $\mathcal{S}_M$ is the permutation group of $M$ elements and the S-matrices $\mathbb{S}_\sigma$ are defined by the decomposition of the permutation into two-cycles with the S-matrix corresponding to the two-cycle $(ij)$ given by
\begin{align}
&\mathbb{S}_{ij} = \frac{u_i- u_j -i}{u_i- u_j + i}.
\end{align}
For instance, $\mathbb{S}_{321} = \mathbb{S}_{23}\mathbb{S}_{13}\mathbb{S}_{12} $. Moreover, $\mathcal{N}$ can be 
expressed in terms of the S-matrices by considering the reflection permutation $\mathcal{N}= 1/\sqrt{\mathbb{S}_{M,M-1,\ldots 1}}$. 
The Bethe state \eqref{eq:BSsu2} is an eigenstate of the spin chain Hamiltonian provided that the rapidities satisfy the Bethe equations
\begin{align}\label{eq:SU2BAE}
1= \Big( \frac{u_n - \ihalf}{u_n + \ihalf} \Big)^L \prod^M_{m\neq n} \frac{u_n - u_m + i }{u_n - u_m - i} .
\end{align}
The flipped spins constitute excitations which propagate along the spin chain with momentum $p$ given by
$u=\frac{1}{2}\cot\left(\frac{p}{2}\right)$. In order that a Bethe eigenstate can be identified
with a single trace operator it must fulfill the cyclicity (i.e. zero-momentum) constraint.
\begin{equation}\label{cyclicity}
\prod_{n=1}^M\Big( \frac{u_n - \ihalf}{u_n + \ihalf} \Big)=1.
\end{equation}
In the $S\!U(2)$ sector the the matrix product
state which implements the insertion of the vevs corresponding to $Z$ and $Y$ into a conformal single trace operator represented by a Bethe eigenstate takes the form~\cite{deLeeuw:2015hxa}
\begin{equation}\label{MPS2}
\left\langle {\rm MPS_k}\,\right|=\mathop{\mathrm{tr}}\nolimits_a
 \prod_{l=1}^{L}\left(\left\langle e_1\right|\otimes t_1^{(k)}
 +\left\langle e_2 \right|\otimes t_2^{(k)}\right),
\end{equation}
where the subscript $a$ refers to the auxiliary $k$-dimensional space in which the generators $t_i$ act. 

In \cite{deLeeuw:2015hxa,Buhl-Mortensen:2015gfd}  it was shown that only states with $L$ and $M$ even and with paired rapidities 
$\{u_i\} = \{-u_i\}$ have a non-trivial overlap with the matrix product state and hence non-vanishing one-point functions.
In particular, for $k=2$, the one-point function can be expressed as%
\begin{equation}\label{C-overlap}
 C_2 
  = 2^{1-L}\sqrt{
  \left[
 \prod_{j}^{}\frac{u_j^2+\frac{1}{4}}{u_j^2}\right]\,\frac{\det G^+}{\det G^-}},
\end{equation} 
where $G^\pm$ are $\frac{M}{2}\times \frac{M}{2}$ matrices with matrix elements:
\begin{equation}
 G^\pm_{jk}=\left(\frac{L}{u_j^2+\frac{1}{4}}-\sum_{n}^{}K^+_{jn}\right)\delta _{jk}
 +K^\pm_{jk},
\end{equation}
and $K^\pm_{jk}$ are defined as
\begin{equation}
 K^\pm_{jk}=\frac{2}{1+\left(u_j-u_k\right)^2}\pm
 \frac{2}{1+\left(u_j+u_k\right)^2}\, .
\end{equation}
This result can be proven from the result for the overlap between a Bethe eigenstate and the $(2m)$-fold raised N\'{e}el state~\cite{Brockmann2014a,Brockmann2014,Brockmann2014b}.  For an alternative derivation, see \cite{Foda:2015nfk}. 
The expression for $C_k$ for arbitrary $k$ was derived in \cite{Buhl-Mortensen:2015gfd}.

A key point in the generalization of the determinant formula~({\ref{C-overlap}) to the $S\!U(3)$ case is the observation, already
made in~\cite{Brockmann2014a} in connection with the study of the overlap between the N\'{e}el state and a Bethe eigenstate, that the matrices $G^{\pm}$ are closely related to  quantities which appear in the Gaudin formula for the
norm of the Bethe eigenstate. 

Consider the $S\!U(2)$ Bethe equations \eqref{eq:SU2BAE} and define the norm functions $\phi_n$ as the logarithm of the right hand side of the Bethe equations
\begin{align}
\phi_n := -i \log\left[ \Big( \frac{u_n - \ihalf}{u_n + \ihalf} \Big)^L \prod^M_{m\neq n} \frac{u_n - u_m + i }{u_n - u_m - i} \right].
\end{align}
The norm of a Bethe state is then completely given in terms of the derivatives of the norm function in the following way
\begin{align}\label{eq:Gaudin}
\langle \{u_i\} | \{u_i\} \rangle = \prod_{i=1}^M \Big[ u_i^2 + \sfrac{1}{4}\Big] \det_{M\times M} \partial_m \phi_n,
\end{align}
where $\partial_m = \partial / \partial u_m$.

Recall that the only non-trivial one-point functions in the $S\!U(2)$ sector are obtained for  Bethe eigenstates with an even number $M$ of rapidities such that $\{u_i\} = \{-u_i\}$. This property of the root set  causes
the Gaudin norm \eqref{eq:Gaudin} to factorize. In order to see this, let us order the Bethe roots as follows $\{u_1,\ldots, u_{\sfrac{M}{2}}, -u_1,\ldots, -u_{\sfrac{M}{2}} \}$. For the corresponding 
eigenstate, the norm matrix $\partial_m \phi_n$ then takes the following form
\begin{equation}
\partial_m \phi_n = 
\begin{pmatrix}
A_1 & A_2 \\
A_2 & A_1
\end{pmatrix}, 
\end{equation}
where
\begin{equation}
A_1 = \left( \partial_m \phi_n \right)_{m,n = 1,\ldots, \sfrac{M}{2}},\hspace*{0.5cm}
A_2 = ( \partial_{m+M/2} \phi_n )_{m,n = 1,\ldots, \sfrac{M}{2}}.
\end{equation}
It is now easy to see that the determinant factorizes 
\begin{align}
&\det \partial_m \phi_n = \det ( A_+)  \det(A_-),
&& A_\pm :=  A_1 \pm A_2.
\end{align}
Remarkably, it turns out that the inner product of the MPS with the Bethe state is proportional to one of these factors
\begin{align}
\langle \mathrm{MPS} | \{u_i\}\rangle \propto \det A_+.
\end{align}
in such a way that the one-point function can be written as
\begin{align}\label{eq:SU2quotient}
C_2 = 2^{1-L}\sqrt{\left[\prod_{m=1}^{M/2} \frac{u_m^2 + \sfrac{1}{4}}{u_m^2}\right] \frac{\det  A_+}{\det  A_-}}.
\end{align}

In other words, we find that for $k=2$, the one-point function $C_2$ is given by the quotient of the factors of the Gaudin determinant, with a slightly modified prefactor. We will see that this structure persists for the $S\!U(3)$ sector as well. 

\section{The $SU(3)$ case\label{SU(3)}}

In order to extend the $S\!U(2)$ sector discussed above to $S\!U(3)$ we extend our definition of the complex scalar fields
as follows
\begin{align}
&Z=\Phi _1+i\Phi _4,
&&Y=\Phi _2+i\Phi _5,
&&& W= \Phi_3+i \Phi_6.
\end{align}
Conformal single trace operators built from these three complex fields can be identified with the Bethe eigenstates of
the integrable $S\!U(3)$ Heisenberg spin chain. The Hamiltonian of the spin chain takes the same form as before, cf.\
eqn.~(\ref{Hamiltonian}), but this time there is a Hilbert space $\mathbb{C}^3$ with basis elements $|e_{1,2,3}\rangle$
associated to each site of the chain.
We now choose the vacuum of the spin chain as $|0\rangle_L := |e_1\ldots e_1\rangle_L$ and map this state to the operator $\mbox{Tr}\, Z^L$.
 Let us also recall the matrix unities $E^i_j : = |e_j \rangle \langle e_i|$. The Bethe Ansatz for the $S\!U(3)$ spin chain is worked out in detail f.inst.\ in \cite{Escobedo:2012ama}. Here, we will only collect the  formulas that will be of importance for the following. The Bethe states are labelled by three discrete parameters $L,M,N$ that correspond to the length of the spin chain, and  the two Dynkin labels of $\alg{su}(3)$ and are constrained to obey $L\geq 2M\geq 4N$.
 In the language of single trace operators, $L$ is again the total number of fields in the operator, $M$ is the number of excitations, 
 i.e.\ $Y$- and $W$-fields and $N$ is the number of $W$-fields. 

A Bethe state will be a linear combination of the form
\begin{align}
&|\{v_i;w_i\}\rangle :=  \mathcal{N} \tilde{\mathcal{N}}\cdot  \nonumber \\
& \sum_{1\leq m_1<\ldots <m_M \leq L}
\sum_{1\leq n_1<\ldots <n_N \leq M}
\sum_{\sigma\in \mathcal{S}_M}
\sum_{\tau\in \mathcal{S}_N} 
\psi_{\sigma\tau} \prod_{r=1}^N (E^2_3)_{m_{n_r}} \prod_{s=1}^M  (E^1_2)_{m_s} |0\rangle,
\end{align}
where $\mathcal{S}_N$ are all permutations of $N$ elements and the coefficient $\psi$ is given by 
\begin{align}
&\psi_{\sigma\tau} := \tilde{\psi}_{\sigma\tau} \, \mathbb{S}_\sigma \prod_{r=1}^M \left[\frac{v_{\sigma_r}+\ihalf}{v_{\sigma_r}-\ihalf}\right]^{m_r} , \\
& \tilde{\psi}_{\sigma\tau} := \tilde{\mathbb{S}}_\tau \prod_{s=1}^N \prod_{r=1}^M \frac{(w_{\tau_s}-v_{\sigma_r}+\ihalf)^{1-\delta_{r,M}}}{w_{\tau_s}-v_{\sigma_r}-\ihalf}.
\end{align}
The factors $\mathcal{N},\tilde{\mathcal{N}}$ again simply correspond to normalizations. Due to the fact that $S\!U(3)$ has rank two, we have two S-matrices $\mathbb{S}_\sigma,\tilde{\mathbb{S}}_\tau$, which are defined by the decomposition of the permutation into two-cycles. The S-matrix corresponding to a single two-cycle $(ij)$ is given by
\begin{align}
&\mathbb{S}_{ij} = \frac{v_i- v_j -i}{v_i- v_j + i},
&\tilde{\mathbb{S}}_{ij} = \frac{w_i- w_j -i}{w_i - w_j + i}.
\end{align}
In terms of the S-matrices, the normalization constants are given by $\mathcal{N} = 1/\sqrt{\mathbb{S}_{M,M-1,\ldots 1}}$ and $\tilde{\mathcal{N}} = 1/\sqrt{\tilde{\mathbb{S}}_{N,N-1,\ldots 1}}$. The total normalization constant is simply a phase, but \eqref{genericCso6} is sensitive to this. Notice that in the case $N=0$ we recover the Bethe wave function for $S\!U(2)$.

In order for the state $|\{v_m;w_n\} \rangle$ with labels $(L,M,N)$ to be an eigenstate the 
spectral parameters $v,w$ have to fulfill the $S\!U(3)$ Bethe equations%
\begin{align}
1&= \Big( \frac{v_m - \ihalf}{v_m + \ihalf} \Big)^L 
\prod^M_{n\neq m} \frac{v_m - v_n + i }{v_m - v_n - i} 
\prod^N_{n=1} \frac{v_m - w_n - \ihalf }{v_m - w_n - \ihalf} ,\label{eq:BAEsu3a}\\
1&=\prod^M_{m= 1} \frac{w_n - v_m - \ihalf }{w_n - v_m + \ihalf} \prod^N_{m\neq n} \frac{w_n - w_m + i }{w_n - w_m - i}.\label{eq:BAEsu3b}
\end{align}
It can be shown that the cyclicity constraint on the states takes the same form as~(\ref{cyclicity}), just with $u$'s replaced by $v$'s.
 The parameters $v$ are usually called the momentum carrying or physical roots and the parameters $w$ are called auxiliary roots.
 
The norm of  an $S\!U(3)$  Bethe eigenstate can be computed via a generalization of the Gaudin formula \cite{Escobedo:2010xs,Escobedo:2012ama}. To this end, one introduces two norm functions corresponding to the two Bethe equations  \eqref{eq:BAEsu3a} and \eqref{eq:BAEsu3b} 
\begin{align}
\phi^v_m &:= -i \log\left[
\Big( \frac{v_m - \ihalf}{v_m + \ihalf} \Big)^L 
\prod^M_{n\neq m} \frac{v_m - v_n + i }{v_m - v_n - i} 
\prod^N_{n=1} \frac{v_m - w_n - \ihalf }{v_m - w_n - \ihalf} 
\right],\\
\phi^w_n &:= -i \log\left[\prod^M_{m= 1} \frac{w_n - v_m - \ihalf }{w_n - v_m + \ihalf} \prod^N_{m\neq n} \frac{w_n - w_m + i }{w_n - w_m - i} \right].
\end{align}
The norm of a Bethe state is then given by
\begin{align}
\langle \{v_m;w_n\} | \{v_m;w_n\} \rangle = \prod_{i=1}^M \Big[ v_i^2 + \sfrac{1}{4}\Big] \det_{(M+N)\times(M+N)} \partial_I \phi_J,
\end{align}
where the generalized indices $I,J = 1,\ldots M, M+1,\ldots M+N$ run over both the momentum carrying and auxiliary Bethe roots. Notice that the prefactor only includes the momentum carrying roots.

Let us now consider the inner product of a Bethe eigenstate $| \{v_m;w_n\} \rangle$ with the Matrix Product state for $k=2$. In this case, the matrix product state is built of Pauli matrices which fulfill
\begin{align}
&\sigma_i^2 = \frac{1}{4},
&& \{\sigma_i,\sigma_j\} = 0.
\end{align}
This implies that we can write any trace that occurs in the one-point function of an operator corresponding to
a Bethe eigenstate with labels $L,M,N$ as
\begin{align}
\pm\mathrm{tr} [\sigma_1^{L-M}\sigma_2^{M-N}\sigma_3^{N}].
\end{align}
It is easy to see that this is only non-zero in the following two cases
\begin{itemize}
\item $L,M,N$ all even
\item $L,N$ odd and $M$ even
\end{itemize}
Hence, only Bethe states with such labels will have a non-trivial one-point function. Moreover, from the anticommutation relations of the Pauli matrices, it quickly follows that
\begin{align}
\langle \mathrm{MPS} | \{v_i;w_i\} \rangle &= 
2^{1-L} \mathcal{N} \tilde{\mathcal{N}} \, \cdot \nonumber  \\
&\sum_{1\leq m_1<\ldots}
\sum_{1\leq n_1<\ldots } (-1)^{\sum m_i + \sum n_i}
\sum_{\sigma\in \mathcal{S}_M}
\sum_{\tau\in \mathcal{S}_N} 
\psi_{\sigma\tau} .
\end{align}
As in the $S\!U(2)$ case, the MPS is not an eigenstate of the Hamiltonian, but is an eigenstate of the momentum operator $P$ with eigenvalue zero. It is also an eigenstate of the third conserved charge $Q^{(3)} = [\mathcal{H}\otimes1,1\otimes \mathcal{H}]$ with eigenvalue zero~\cite{deLeeuw:2015hxa}. This is analogous to the $S\!U(2)$ case. This shows that only Bethe eigenstates
which are annihilated by the third conserved charge can have non-trivial one-point functions. Such states are parity singlet states 
for which $\{v_i\} = \{-v_i\}$.
This pairing of the physical roots has implications for the auxiliary roots $w$ as well. By combining the Bethe equations \eqref{eq:BAEsu3a} for $\pm v_m$, we derive the following set of equations
\begin{align}
1 = \prod_{n=1}^N \frac{v_m - w_n + \ihalf}{v_m - w_n - \ihalf}\frac{v_m + w_n - \ihalf}{v_m + w_n + \ihalf}.
\end{align}
This defines a polynomial equation of degree $N-1$ in $v^2_m$. So, for fixed $w_n$ this gives us $N-1$ possible solutions. Since Bethe states are trivial if two Bethe roots coincide and since $M\geq2N$, we find that the above equation can not lead to a non-trivial Bethe state unless it is trivially satisfied. This means the auxiliary roots $w_n$ must satisfy $\{w_n\} = \{-w_n\} $. For even $N$ this simply means that the auxiliary roots also must come in pairs. For odd $N$ this means that one of the auxiliary roots is trivial $w_N=0$ and the remaining roots must come in pairs.
In the following we will present a closed determinant formula for the one-point functions of the $S\!U(3)$ sector, similar to the one for the $S\!U(2)$ subsector. We have to distinguish two cases, namely $N$ even and $N$ odd.

For even $N$, both the momentum carrying and the auxiliary roots come in pairs. Let us order them as $\{v_1,\ldots, v_{\frac{M}{2}},-v_1,\ldots, - v_{\frac{M}{2}}\}$ and $\{w_1,\ldots, w_{\frac{N}{2}},-w_1,\ldots, - w_{\frac{N}{2}}\}$. It is again easy to see that the norm matrix $\partial_I \phi_J$ takes a very symmetric form
\begin{align}
\partial_I \phi_J = 
\begin{pmatrix}
A_1 & A_2 & B_1 & B_2 \\
A_2 & A_1 & B_2 & B_1 \\
B^t_1 & B^t_2 & C_1 & C_2 \\
B^t_2 & B^t_1 & C_2 & C_1
\end{pmatrix},
\end{align}
where $t$ stands for transposition and 
\begin{align}
&A_1 = \partial_{v_m} \phi^v_n, \qquad m=1,\ldots \sfrac{M}{2},\, n=1,\ldots \sfrac{M}{2} \\ 
&A_2 = \partial_{v_m} \phi^v_n, \qquad m=\sfrac{M}{2}+1,\ldots M,\, n=1,\ldots \sfrac{M}{2}\\
&B_1 = \partial_{v_m} \phi^w_n, \qquad m=1,\ldots \sfrac{M}{2},\, n=1,\ldots \sfrac{N}{2} \\ 
&B_2 = \partial_{v_m} \phi^w_n, \qquad m=\sfrac{M}{2}+1,\ldots M,\, n=1,\ldots \sfrac{N}{2}\\
&C_1 = \partial_{w_m} \phi^w_n, \qquad m=1,\ldots \sfrac{N}{2},\, n=1,\ldots \sfrac{N}{2} \\ 
&C_2 = \partial_{w_m} \phi^w_n, \qquad m=\sfrac{N}{2}+1,\ldots N,\, n=1,\ldots \sfrac{N}{2}.
\end{align}
The determinant of the norm matrix factorizes. In particular, we have
\begin{align}
&\det \partial_I \phi_J = \det G_+ \det G_-, 
&& G_\pm = 
\begin{pmatrix}
A_\pm & B_\pm \\
B^t_\pm & C_\pm
\end{pmatrix},
\end{align}
where $A_\pm = A_1 \pm A_2$ etc. 
This suggests a  direct generalization of \eqref{eq:SU2quotient}, more precisely
\begin{align}\label{eq:Neven}
C(\{v_i;w_i\}) = 2^{1-L}\sqrt{\left[\prod_{m=1}^{M/2} \frac{v_m^2+ \sfrac{1}{4}}{v_m^2}\right]\left[\prod_{n=1}^{N/2} \frac{w_n^2+ \sfrac{1}{4}}{w_n^2}\right] \frac{\det G_+}{\det G_-}}.
\end{align}
For $N=0$ this reduces to the $S\!U(2)$ result. Moreover, for $N=2$ we checked the overlap formula numerically for Bethe states with labels $(L,M,N) = (8,4,2),(10,4,2),(12,6,2)$ for  different numerical solutions of the Bethe equations and perfect agreement is found to 50 digits. The next different value of $N$ would be $N=4$ and since $L\geq 2M\geq4N$ we find that the smallest such state has labels $(L,M,N)=(16,8,4)$. This Bethe state has in the order of $10^{12}$ terms, which makes it unaccessible from a practical point of view.

For odd $N$, the Bethe roots take the form 
$$
\{v_1,\ldots, v_{\frac{M}{2}},-v_1,\ldots, - v_{\frac{M}{2}}\} \cup \{w_1,\ldots, w_{\frac{N-1}{2}},-w_1,\ldots, - w_{\frac{N-1}{2}},0\}.
$$
 Again, the norm matrix $\partial_I \phi_J$ shows a symmetric form, but the additional trivial auxiliary root $w_N = 0$ introduces an additional row and column. More precisely, we have
\begin{align}
\partial_I \phi_J = 
\begin{pmatrix}
A_1 & A_2 & B_1 & B_2 & D_1 \\
A_2 & A_1 & B_2 & B_1 & D_1\\
B^t_1 & B^t_2 & C_1 & C_2 & D_2\\
B^t_2 & B^t_1 & C_2 & C_1 & D_2 \\ 
D^t_1 & D^t_1 & D^t_2 & D^t_2 & D_3 
\end{pmatrix},
\end{align}
where again $t$ stands for transposition and 
\begin{align}
&A_1 = \partial_{v_m} \phi^v_n, \qquad m=1,\ldots \sfrac{M}{2},\, n=1,\ldots \sfrac{M}{2} \\ 
&A_2 = \partial_{v_m} \phi^v_n, \qquad m=\sfrac{M}{2}+1,\ldots M,\, n=1,\ldots \sfrac{M}{2}\\
&B_1 = \partial_{v_m} \phi^w_n, \qquad m=1,\ldots \sfrac{M}{2},\, n=1,\ldots \sfrac{N-1}{2} \\ 
&B_2 = \partial_{v_m} \phi^w_n, \qquad m=\sfrac{M}{2}+1,\ldots M,\, n=1,\ldots \sfrac{N-1}{2}\\
&C_1 = \partial_{w_m} \phi^w_n, \qquad m=1,\ldots \sfrac{N-1}{2},\, n=1,\ldots \sfrac{N-1}{2} \\ 
&C_2 = \partial_{w_m} \phi^w_n, \qquad m=\sfrac{N-1}{2}+1,\ldots N,\, n=1,\ldots \sfrac{N-1}{2}\\
& D_1 = \partial_{w_N} \phi^v_m, \qquad m=1,\ldots \sfrac{M}{2}\\ 
& D_2 = \partial_{w_N} \phi^W_n, \qquad n=1,\ldots \sfrac{N-1}{2}\\ 
& D_3 = \partial_{w_N} \phi^W_N.
\end{align}
Remarkably, the determinant of the norm matrix again factorizes
\begin{align}
&\det \partial_I \phi_J = \det G_+ \det G_-, \\
& G_+ = 
\begin{pmatrix}
A_+ & B_+ \\
B^t_+ & C_+
\end{pmatrix}, \,\,\,\,
 G_- = 
\begin{pmatrix}
A_- & B_- & D_1\\
B^t_- & C_- & D_2 \\
2 D^t_1 & 2D^t_2 & D_3 
\end{pmatrix}.
\end{align}
suggesting the following formula for the one-point functions
\begin{align}\label{eq:Nodd}
C(\{v_i;w_i\}) = 2^{1-L}\sqrt{\left[\prod_{m=1}^{M/2} \frac{v_m^2+ \sfrac{1}{4}}{v_m^2}\right]\left[\prod_{n=1}^{(N-1)/2} \frac{w_n^2+ \sfrac{1}{4}}{w_n^2}\right] \frac{1}{4}\frac{\det G_+}{\det G_-}}.
\end{align}
Comparing \eqref{eq:Neven} to \eqref{eq:Nodd}, we see that they are exactly the same up to the special terms that involve $w_N$. Since this Bethe root is zero, it does not appear in the product that multiplies the quotient of determinants, but rather gives an additional factor of $\sfrac{1}{4}$. We also find that $G_-$ is modified by supplementing it with an additional row and column. 
The formula~(\ref{eq:Nodd}) is supported by a number of highly non-trivial checks. We have checked our results analytically for $N=1, M=2,4$ \textit{for any} $L$. For $N=3$ we checked the overlap formula numerically for Bethe states with labels $(L,M,N) =(13,6,3)$. This state has almost 150 million terms, which all conspire to give a perfect agreement for up to 50 digits.

\section{Discussion and Conclusion\label{Conclusion}}

Although our various checks of the formulas for the one-point functions of the $S\!U(3)$ sub-sector give ample support that
the formulas are correct, an analytical proof thereof would be highly desirable. What  is missing is a proof that the overlap
of a Bethe eigenstate with the matrix product state can be written as a specific determinant in the same way as it was the
case for the $S\!U(2)$ sub-sector.  In the $S\!U(2)$ case this could be proved by showing that the matrix product state was cohomologically
equivalent to the N\'{e}el state in the case of $M=L/2$~\cite{deLeeuw:2015hxa} and to an appropriately raised 
N\'{e}el state for $M<L/2$~\cite{Buhl-Mortensen:2015gfd} and combining this with the  the fact that the inner product between a Bethe eigenstate and the latter states were known 
already~\cite{Brockmann2014b,Brockmann2014,Brockmann2014a}.
Alternatively, the overlaps for $M<L/2$ could be related to overlaps involving so-called partial N\'{e}el states~\cite{deLeeuw:2015hxa} and evaluated by identifying them with a partial version of the reflecting boundary domain wall 
partition function of the six vertex model~\cite{Foda:2015nfk}.  Investigating the concept of N\'{e}el states and their
overlaps with Bethe eigenstates in case of the $S\!U(3)$ spin chain would be interesting not only for one-point
functions in the present defect version of ${\cal N}=4$ SYM, but also in its own right.

A question closely related to the above is whether the other determinant in the product formula for the Gaudin norm can
be understood as the inner product of a Bethe eigenstate with some other state. This question also lacks an answer in the case of the $S\!U(2)$ spin chain.

The results of the present paper deal entirely with the case $k=2$ where $k$ is equal to the difference in the rank of the gauge group  between the two sides of the defect in the dCFT set-up and equal to the amount of background gauge field flux in the dual string theory set-up. At the moment comparisons between gauge theory and string theory are possible only in a certain double scaling limit which implies that one should consider  $k\rightarrow \infty$~\cite{Nagasaki:2012re}. Hence, it would be very
interesting to extend the present results to the case of general $k$. In the case of $S\!U(2)$ it was possible to derive a two-step
recursion relation which related the one-point function for higher even values of $k$ to that for $k=2$ and the 
one-point function for all higher odd values of $k$ to that for $k=3$. Solving the recursion relation furthermore revealed that
the one-point function for any $k$ could be written as a pre-factor times that for $k=2$~\cite{Buhl-Mortensen:2015gfd}.
We find that this is still the case for $N=1$ with the same prefactor and a rescaling $L\rightarrow L+1$ but our analysis
has not revealed any similar structure of the result in other cases.  The above mentioned recursion relation followed from
the fact that the result of acting with the $S\!U(2)$ transfer matrix on the matrix product state corresponding to a representation of rank
$k$ could be decomposed into matrix product states corresponding to representations of rank $k-2$  and $k-4$, respectively.  In the present case we have not been able to identify such a decomposition.

 The fact that  the calculation of one-point functions entails the natural appearance of an auxiliary vector space via the matrix product state might be taken as an indication that a reformulation of the calculation in the language of the algebraic Bethe ansatz could be possible and maybe even   advantageous.  Developing such a reformulation constitutes an interesting future
direction of investigation. It would likewise be interesting to  extend our discussion to the full $S\!O(6)$ scalar sector of $\mathcal{N}=4$ SYM.

 \vspace{0.5cm}
 
\section*{Acknowledgments}
M.d.L. and C.K.\ were supported by FNU through grants number DFF -- 1323 -- 00082 and DFF-4002-00037. Stefano
Mori was supported by "Fondazione Angelo Della Riccia''.

\vspace*{0.3cm}



 \bibliographystyle{elsarticle-num} 
 \bibliography{letter.bib}

\vspace*{0.5cm}


\end{document}